\documentclass[prd,twocolumn,amsmath,amsfonts,amssymb,showpacs]{revtex4} 
\usepackage{epsfig,psfrag,graphicx,bm,color}
\usepackage{float}

\begin{document}
\title{Inert Doublet Model and LEP II Limits}

\author{Erik Lundstr\"om}
\email{erik@physto.se}
\affiliation{{D}epartment of Physics, Stockholm University, AlbaNova
 University Center, SE - 106 91 Stockholm, Sweden}
\author{Michael Gustafsson}
\email{gustafss@pd.infn.it} \affiliation{INFN, Sezione di Padova,
Department of Physics ``Galileo Galilei'', Via Marzolo 8, I-35131,
Padua, Italy} \affiliation{Department of Physics, Stockholm
University, AlbaNova University Center, SE - 106 91 Stockholm,
Sweden}
\author{Joakim Edsj\"o}
\email{edsjo@physto.se}
\affiliation{Department of Physics, Stockholm University, AlbaNova
 University Center, SE - 106 91 Stockholm, Sweden}

\begin{abstract}
The inert doublet model is a minimal extension of the standard model
introducing an additional SU(2) doublet with new scalar particles
that could be produced at accelerators. While there exists no LEP II
analysis dedicated for these inert scalars, the absence of a signal
within searches for supersymmetric neutralinos can be used to
constrain the inert doublet model. This translation however requires
some care because of the different properties of the inert scalars
and the neutralinos. We investigate what restrictions an existing
DELPHI collaboration study of neutralino pair production can put on
the inert scalars and discuss the result in connection with dark
matter. We find that although an important part of the inert doublet
model parameter space can be excluded by the LEP II data, the
lightest inert particle still constitutes a valid dark matter
candidate.
\end{abstract}

\maketitle

\newcommand{\nn}{\nonumber}
\newcommand{\ga}{\gamma}
\hyphenation{micrOMEGAs higg-sino}

\section{Introduction}

The inert doublet model (IDM) extension of the particle standard
model has recently attracted attention within the particle physics
and cosmology community. Despite its simplicity - a two Higgs
doublet model with an imposed unbroken $Z_2$ parity - it displays a
remarkably rich phenomenology. After being touched upon already in
the 1970s \cite{Deshpande:1977rw} the model has in later years been
studied within various contexts such as light neutrino generation
\cite{Ma:2006km,Ma:2006fn}, leptogenesis \cite{Ma:2006fn}, Higgs
phenomenology \cite{Barbieri:2006dq,Cao:2007rm}, improved
naturalness \cite{Barbieri:2006dq,Casas:2006bd}, electroweak
symmetry breaking \cite{Hambye:2007vf}, and as a dark matter
candidate
\cite{Ma:2006km,Ma:2006fn,Barbieri:2006dq,LopezHonorez:2006gr,Gustafsson:2007pc}.

Besides the standard model-like Higgs boson, $h$, the IDM contains
three additional scalar fields: two neutral, $H^0$ and $A^0$, and
one charged, $H^{\pm}$. Because of the exact $Z_2$ parity none of
the new particles have direct (Yukawa-) couplings to fermions and
are hence called \textit{inert} scalars (or sometimes even inert
Higgs bosons \cite{Barbieri:2006dq,Gustafsson:2007pc}). Because the
new doublet still couples directly to $h$ and the gauge bosons (and
does not generate standard model masses) some have preferred to
instead denote the model the dark scalar doublet model
\cite{Cao:2007rm}.

The $Z_2$ parity also ensures the stability of the lightest inert
scalar, which hence, if it is either $H^0$ or $A^0$, constitutes a
good dark matter candidate belonging to the class of weakly
interacting massive particles (WIMPs). For masses below 80 GeV it
can leave a relic abundance in agreement with the WMAP constraints
\cite{Spergel:2006hy,Komatsu:2008hk} and hence explain all the dark
matter \cite{Barbieri:2006dq,LopezHonorez:2006gr,Gustafsson:2007pc}.

Another striking property of the IDM is that it can allow for a
heavy ($\sim\! 500$ GeV) standard model Higgs boson $h$ without
violating data from electroweak precision tests
\cite{Barbieri:2006dq}. Although the direct Higgs searches are not
in conflict with the 95\% C.L. upper Higgs mass bound set by the
precision tests (assuming only standard model physics), they are
well above the indicated central value \cite{lepewwg}. If upcoming
accelerator searches fail to detect a light ($\lesssim 200$ GeV)
Higgs boson, the IDM may constitute an attractive explanation.

Although the IDM (in contrast to \textit{e.g.}~supersymmetric
models) seems to lack any deeper fundamental motivations, its
strength instead lies in its simplicity, and the model can be
regarded as an archetype for more comprehensive extensions of the
standard model (like \cite{Pierce:2007ut} or \cite{Ma:2007kt}). Just
like generic spin-1/2 and spin-1 WIMPs could be represented by the
supersymmetric neutralino and the first Kaluza-Klein excitation of
the photon, respectively, the lightest inert particle could be used
as a representative of spin-0 WIMP dark matter candidates.

Experimental bounds and signatures of the IDM have been investigated
in several papers. Studies show that the WIMP-nucleon scattering
cross section in general lies below the sensitivity of current deep
underground direct detection experiments
\cite{Barbieri:2006dq,LopezHonorez:2006gr}. The prospects for
indirect detection via gamma-rays produced in dark matter
self-annihilation processes in the galactic halo look more promising
\cite{LopezHonorez:2006gr,Gustafsson:2007pc,Serpico:2008ga},
especially since inert scalar dark matter has the ability to produce
extraordinarily clear spectral lines \cite{Gustafsson:2007pc}. LEP I
data on the width of the $Z$ boson force the sum of the $H^0$ and
$A^0$ masses to be larger than the $Z$ mass
\cite{Gustafsson:2007pc,Cao:2007rm}, while the LEP electroweak
precision tests put constraints on the inert scalar mass splittings
as a function of the $h$ mass \cite{Barbieri:2006dq}. Moreover,
detection at the upcoming LHC seems feasible \cite{Cao:2007rm}.

What has been missing so far in the investigation of the IDM is a
closer study in connection with the existing LEP II data. So far the
LEP II limits on the IDM are only rough estimates based on direct
usage of production cross section limits from existing analyses of
the minimal supersymmetric standard model (MSSM)
\cite{Barbieri:2006dq,Pierce:2007ut}. Although this may be suitable
as a first approximation, there are certainly some important
differences between the IDM and the MSSM which threaten to alter the
conclusions.

In this paper we take a closer look at the limits LEP II
data can set on the IDM.
We make use of an existing DELPHI collaboration study of neutralino
pair production \cite{EspiritoSanto:2003} where detailed signals and
backgrounds have already been simulated. By generating
$H^0A^0$ events and mimicking the cuts performed on the
dataset in \cite{EspiritoSanto:2003}, we can estimate upper limits on the
corresponding production cross sections. These limits are then used
to constrain the masses of the inert scalars.

The outline of the paper is as follows. In Section II we present the
IDM setup in more detail and discuss its theoretical and
experimental constraints. In Section III we motivate the need for a
more careful LEP II analysis, and introduce the method used here to
constrain the IDM parameter space. In Section IV we present the
results, and discuss those in connection with dark matter. Section V
includes some final remarks and conclusions. A short summary of this
paper is given in Section VI. Appendices A and B provide detailed
lists of the cuts used for the acoplanar jets and acoplanar leptons
selections, respectively.

\section{The Inert Doublet Model}

\subsection{Description}
Although electroweak symmetry breaking can be achieved with a single
Higgs doublet, a more complicated Higgs sector is not excluded.
Since problems like dark matter and neutrino mass generation seem to
require new physics beyond the standard model, minimal extensions
could be considered as attractive models compatible with the
principle of Occam's razor.

A simple and well known extension of the standard model Higgs sector
is the two Higgs doublet model (THDM), in which there exist two
scalar doublets, $H_1$ and $H_2$. In order not to be in conflict
with the tight constraints on the magnitude of flavor changing
neutral currents a $Z_2$ parity, under which $H_1\!\rightarrow\!
H_1$ and $H_2\!\rightarrow\! -H_2$, is imposed. The most general
renormalizable and CP conserving potential of such a model is
\begin{eqnarray} \label{eq:potential}
V\!\!&=&\!\!\mu_1^2 \vert H_1\vert^2 + \mu_2^2 \vert H_2\vert^2
+ \lambda_1 \vert H_1\vert^4 + \lambda_2 \vert H_2\vert^4\cr
 &+&\!\! \lambda_3 \vert H_1\vert^2 \vert H_2 \vert^2
 + \lambda_4 \vert H_1^\dagger H_2\vert^2
 + {\lambda_5} Re\!\left[(H_1^\dagger H_2)^2\right]\!\!,
\end{eqnarray}
where $\mu^2_i$ and $\lambda_i$ are real parameters. Whether the
Higgs doublets develop vacuum expectation values (vevs) or not of
course depends on the values of the parameters in the potential
\cite{Deshpande:1977rw}. Even if one usually assumes that both
doublets develop vevs, there is \textit{a priori} nothing which
requires that phase.

The IDM is a THDM where only one of the Higgs doublets, $H_1$,
acquires a vev, $v$. Hence, $H_1$ closely corresponds to the
ordinary standard model Higgs doublet. Moreover, the IDM belongs to
the class of type-I THDMs in which all standard model fields are
taken to be $Z_2$ even. As a consequence no Yukawa terms including
$H_2$ are allowed by the symmetry. The nonexistence of a vev, $v_2$,
for $H_2$ leaves the imposed $Z_2$ parity unbroken and ensures the
absence of mixing between the components of $H_1$ and those of
$H_2$. Thus, the fields belonging to $H_2$ are inert in the sense
that they do not couple directly to fermions, and the lightest of
them is automatically stable. It should be noted that the IDM is not
the $v_2\!\rightarrow\! 0$ limit of a THDM in which both Higgs
doublets develop vevs.

After giving masses to the gauge bosons, $H_1$ has one physical
degree of freedom left: the real scalar field $h$. Since $h$ closely
resembles the Higgs particle of the standard model it will here be
called the standard model Higgs boson. In addition, $H_2$ includes
the neutral CP-even $H^0$, the neutral CP-odd $A^0$, and the charged
$H^{\pm}$ inert scalars. The masses of the particles are (at tree
level) given by
\begin{eqnarray} \label{eq:masses}
    m_{h}^2   &=& -2 \mu_1^2,\cr
  m_{H^0}^2 &=& \mu_2^2 + (\lambda_3 + \lambda_4 + \lambda_5) v^2,\cr
  m_{A^0}^2 &=& \mu_2^2 + (\lambda_3 + \lambda_4 - \lambda_5) v^2,\cr
  m_{H^\pm}^2 &=& \mu_2^2 +  \lambda_3 v^2,
\end{eqnarray}
where $v=m_h/\sqrt{4\lambda_1}$ is the vev of $H_1$. (The measured
gauge boson masses determine $v\approx \mathrm{175}$ GeV.)

If either of the electrically neutral $H^0$ or $A^0$ constitutes
the lightest inert particle, it could be a good WIMP dark matter
candidate. By assuming $m_{H^0}\!<\!m_{A^0}$ we will from now on
have $H^0$ as our potential dark matter particle, although
the roles of $H^0$ and $A^0$ are in general interchangeable.

The mass difference between the $A^0$ and $H^0$ is
frequently appearing in calculations, and for later convenience we
define
\begin{eqnarray}
    \Delta m\equiv m_{A^0}\!-m_{H^0}.
\end{eqnarray}

\subsection{Constraints}
The requirements of vacuum stability and perturbativity set
theoretical constraints on the model. If and only if
\begin{eqnarray} \label{eq:vacuumstability}
    \lambda_{1,2}&>&0 \cr
    \lambda_3,\lambda_3+\lambda_4-\left|\lambda_5\right|&>&-2\sqrt{\lambda_1 \lambda_2}
\end{eqnarray}
the potential $V$ of Eq.~(\ref{eq:potential}) is bounded from below,
ensuring a stable vacuum. We will follow \cite{Barbieri:2006dq} and
adopt \pagebreak
\begin{eqnarray} \label{eq:perturbativity}
    \lambda_3^2+\left(\lambda_3+\lambda_4\right)^2+\lambda_5^2&<&12\lambda_1^2 \cr
    \lambda_2&<&1
\end{eqnarray}
as sufficient conditions for a perturbatively well-behaved model.

Observational constraints come from direct and indirect detection
experiments, measurements of the cosmic microwave background, and
accelerator searches.

The absence of WIMP-nucleon scattering signals in existing direct
detection experiments disfavors models very close to having a
Peccei-Quinn symmetry $m_{H0} = m_{A0}$ \cite{Barbieri:2006dq}.
Other direct detection constraints on the IDM parameter space are
not considered in this work [4,8].
\cite{Barbieri:2006dq,LopezHonorez:2006gr}.

Indirect detection through gamma-rays produced by annihilating $H^0$
pairs in and around the galactic center has been studied in several
papers \cite{LopezHonorez:2006gr,Gustafsson:2007pc,Serpico:2008ga}.
Although IDM models (with $m_{H^0}\!<\!m_W$ and large values of
$m_h$) have been shown capable of producing spectacularly clear
monochromatic line signals \cite{Gustafsson:2007pc}, the absolute
photon fluxes are subject to large astrophysical uncertainties.
Also, the most interesting photon energy range has not yet been
covered (but is currently examined by the Fermi Gamma-ray Space
Telescope \cite{GLAST}). Thus it is not yet possible to derive any
robust constraints from indirect detection.

The WMAP data on the cosmic microwave background radiation limits
the relic abundance of dark matter, $\Omega_{DM}$, to 0.094
$\!<\Omega_{DM} h^2<$\! 0.129 \cite{Spergel:2006hy}, where $h$ is
the Hubble constant in units of 100 km Mpc$^{-1}$s$^{-1}$. The lower
bound need only be respected if the IDM alone is to solve the dark
matter problem. However, the relic density of $H^0$ particles must
always respect the upper bound, since they otherwise would contribute
to too much cold dark matter.

While the electroweak precision tests in the absence of new physics
beyond the standard model favor a light Higgs boson, this does not
remain true once the inert scalars are taken into account. In
\cite{Barbieri:2006dq} it was shown that the inert particles make
important contributions to the electroweak observable $T$ according
to
\begin{eqnarray} \label{eq:DeltaT}
    \Delta T\approx\frac{1}{24\pi^2\alpha  v^2}\left(m_{H^\pm}-m_{A^0}\right)\left(m_{H^\pm}-m_{H^0}\right),
\end{eqnarray}
where $\alpha$ is the fine-structure constant. With appropriate
inert scalar mass splittings the IDM can hence compensate for the
otherwise too small values of $T$ arising for heavy standard model
Higgs bosons. From Fig.~1 in \cite{Barbieri:2006dq} we have
estimated the allowed $\Delta T$ range (at 68\% C.L.) as a function
of $m_h$. Together with Eq.\,(\ref{eq:DeltaT}) this translates into
the constraint
\begin{eqnarray} \label{eq:EWPT}
    \left(m_{H^\pm}-m_{A^0}\right)\left(m_{H^\pm}-m_{H^0}\right)\ \ \ \ \ \ \ \ \ \ \ \ \ \ \ \ \ \ \cr
    \cr
  \approx 24\pi^2\alpha v^2 \left[0.15\ln\left(\frac{m_h}{m_Z}\right)\pm0.1\right].
\end{eqnarray}

As first noted in \cite{Gustafsson:2007pc} and more explicitly
argued in \cite{Cao:2007rm} the precise LEP I measurements of the
$Z$ boson width forbids the decay channel $Z\rightarrow
H^0A^0$. Hence it is required that
\begin{eqnarray} \label{eq:Zwidth}
    m_{H^0}+m_{A^0}>m_Z.
\end{eqnarray}

When it comes to limits set by the LEP II experiments, previous
studies have provided rough estimates based on existing analyses for
neutralinos, $\tilde{\chi}_i^0$, within the MSSM. By simply applying
the same upper limits on the $e^+e^-\rightarrow H^0A^0$ cross
section as those set by the different LEP collaborations on
$\tilde{\chi}_1^0 \tilde{\chi}_2^0$ production, one finds that
the LEP II data seem capable of ruling out important parts of the parameter space \cite{Barbieri:2006dq,Pierce:2007ut}.
These observations bring up the question of what
restrictions a more specific study of $H^0 A^0$ production could put
on the IDM parameter space.

\section{LEP II analysis}

\subsection{Motivation}
Although a complete analysis including optimization of cuts, a
detailed detector simulation and standard model background event
generation would put the most accurate LEP II limits on the IDM, it
also demands a lot of effort to be carried out. On the other hand,
the earlier estimates of LEP II constraints
\cite{Barbieri:2006dq,Pierce:2007ut} are quite coarse and, as
motivated below, need to be improved (or possibly verified). In this
work we have therefore chosen to reuse cuts, detector acceptances,
simulated backgrounds and derived production cross section limits
from an existing MSSM analysis \cite{EspiritoSanto:2003}, and
instead focus on IDM signal event generation and efficiency
determination. This method, which will be described in detail in
Section III B, is careful enough to put more accurate limits on the
$H^0$ and $A^0$ masses than the current existing estimates..

The different LEP experiments have separately searched for
neutralinos via the production process $e^+e^-\rightarrow
\tilde{\chi}_1^0 \tilde{\chi}_2^0$ followed by the decay
$\tilde{\chi}_2^0\rightarrow \tilde{\chi}_1^0 f\bar{f}$
\cite{Acciarri:1999km,Barate:2000tu,Abdallah:2003xe,Abbiendi:2003sc}.
The lightest neutralino, $\tilde{\chi}_1^0$, is assumed to be stable
and would show up as missing energy, while the fermion pair,
$f\bar{f}$, could be detected as acollinear jets or leptons.

\begin{figure}
    \centering
        \includegraphics[width=0.4\textwidth]{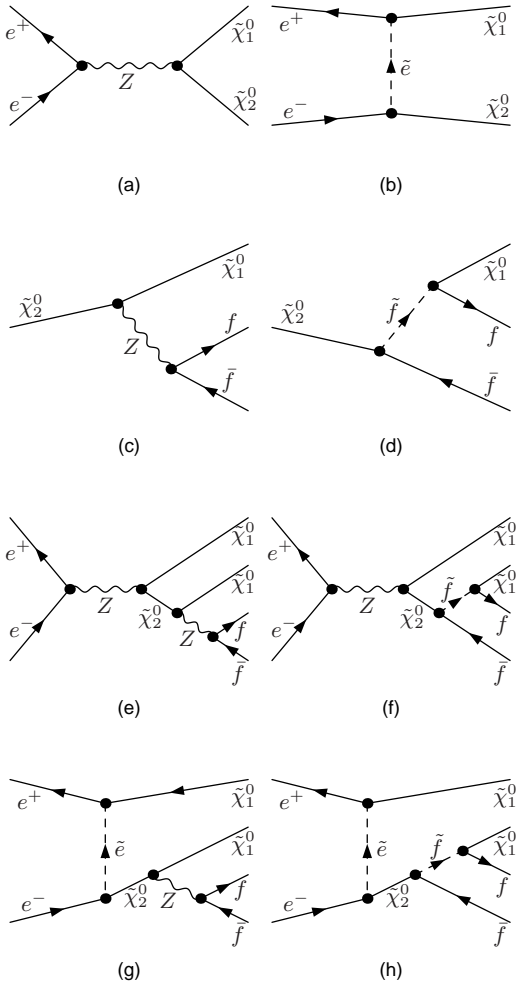}
\caption{Representative Feynman diagrams contributing to
$\tilde{\chi}_1^0 \tilde{\chi}_2^0$ events at LEP. (a)-(d) show the
process factorized into
$e^+e^-\rightarrow\tilde{\chi}_1^0\tilde{\chi}_2^0$ production, (a)
and (b), and subsequent $\tilde{\chi}_2^0\rightarrow\tilde{\chi}_1^0
f\bar{f}$ decay, (c) and (d). (e)-(h) show the unfactorized process
$e^+e^-\rightarrow\tilde{\chi}_1^0\tilde{\chi}_2^0\rightarrow\tilde{\chi}_1^0\tilde{\chi}_1^0f\bar{f}$.}
  \label{fig:susy8}
\end{figure}

\begin{figure}
    \centering
        \includegraphics[width=0.21\textwidth]{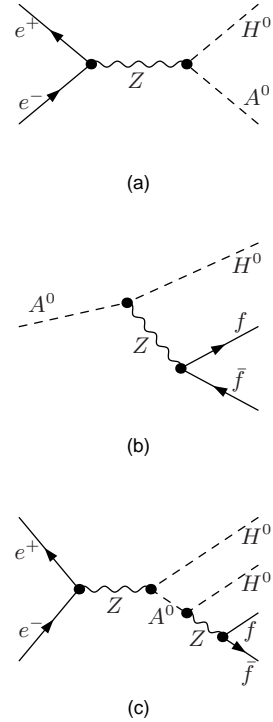}
        \caption{Feynman diagrams contributing to $H^0 A^0$ events at LEP.
(a) and (b) show the process factorized into $e^+e^-\rightarrow H^0
A^0$ production, (a), and subsequent $A^0\rightarrow H^0 f\bar{f}$
decay, (b). (c) shows the unfactorized process $e^+e^-\rightarrow
H^0 A^0\rightarrow H^0 H^0 f\bar{f}$.}
    \label{fig:idm3}
\end{figure}

As noted in \cite{Cao:2007rm}, there are several reasons why the
limits derived for $\tilde{\chi}_1^0 \tilde{\chi}_2^0$ production
need not apply to $H^0 A^0$ production. The inert particles are
scalars while the neutralinos carry spin 1/2, and in principle spin
correlation effects could make a difference. The production of $H^0
A^0$ pairs only proceeds via s-channel spin-1 $Z$ bosons,
$e^+e^-\rightarrow Z\rightarrow H^0 A^0$, forcing the spinless
outgoing scalars into states with large transverse momentum, while
neutralinos have the ability to conserve angular momentum with the
help of their intrinsic spin. Moreover, neutralino production
differs in that it can also take place via t-channel selectron
exchange. Similarly, $A^0$ only decays according to $A^0\rightarrow
H^0 Z\rightarrow H^0 f\bar{f}$, while both
$\tilde{\chi}_2^0\rightarrow \tilde{\chi}_1^0
Z\rightarrow\tilde{\chi}_1^0 f\bar{f}$ and
$\tilde{\chi}_2^0\rightarrow\tilde{f}\bar{f}\rightarrow\tilde{\chi}_1^0
f\bar{f}$ are allowed (where $\tilde{f}$ denotes a sfermion).
Figs.~\ref{fig:susy8}(a)-(d) and Figs.~\ref{fig:idm3}(a)-(b) show
Feynman diagrams contributing to the production and decay processes
for the MSSM and IDM, respectively. In summary, there are many
effects which threaten to spoil a simple usage of the LEP II MSSM
limits directly onto the IDM.

\begin{figure*}[p]
  \includegraphics[height=58mm]{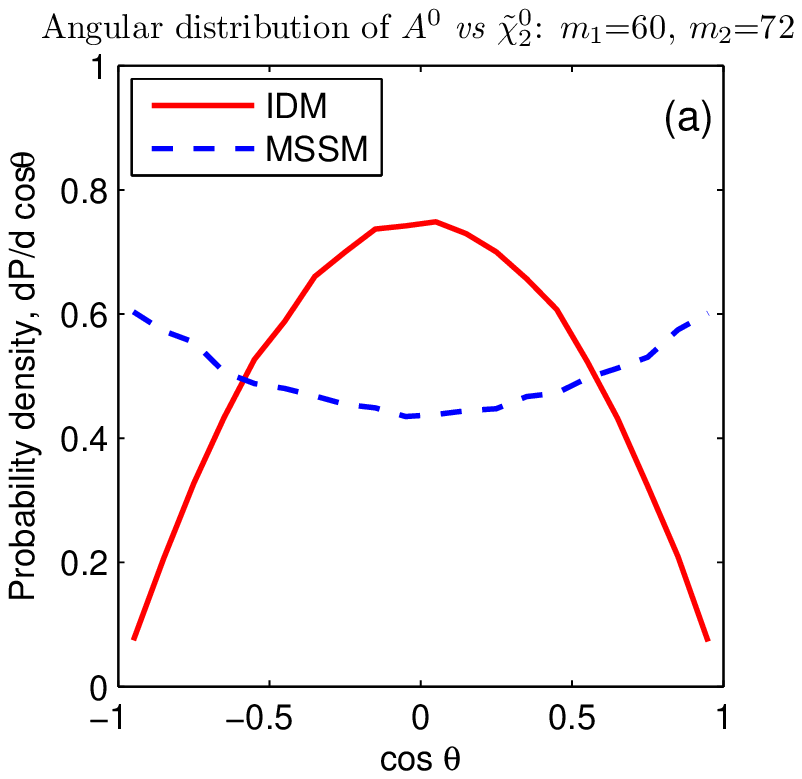}
  \includegraphics[height=58mm]{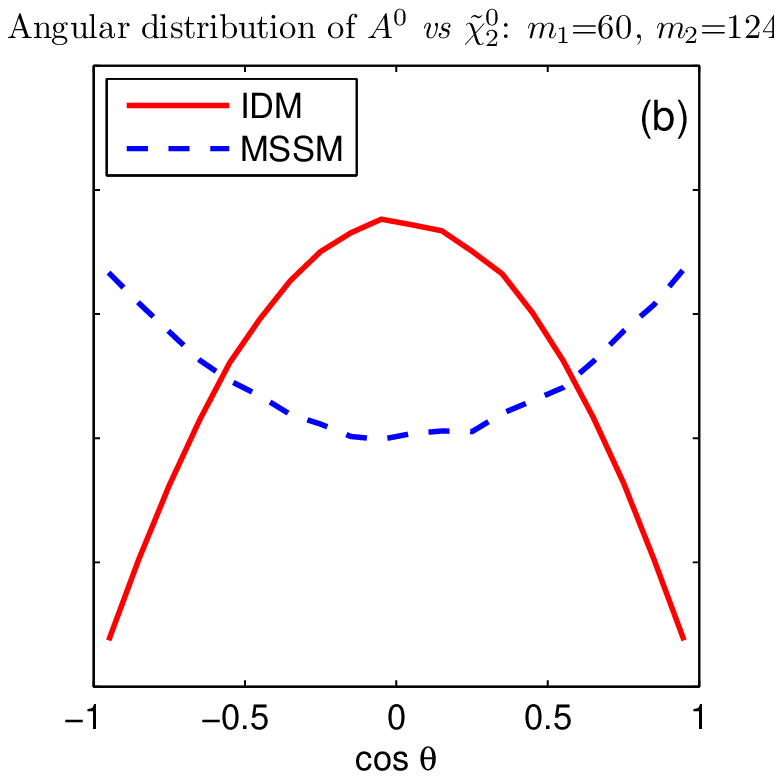}
  \includegraphics[height=58mm]{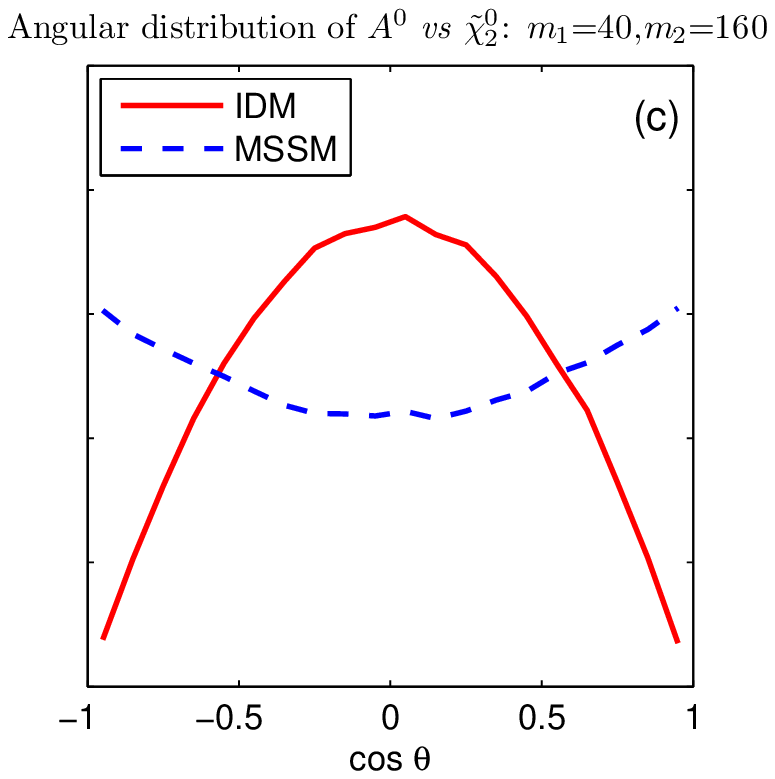}
\caption{Angular distribution of $A^0$ (IDM) \emph{versus}
$\tilde{\chi}_2^0$ (MSSM) produced in $e^+e^-\rightarrow
\tilde{\chi}_1^0 \tilde{\chi}_2^0$ and $e^+e^-\rightarrow H^0 A^0$, respectively. The
header of each subfigure displays (in units of GeV) the mass $m_1$ of $H^0$ and
$\tilde{\chi}_1^0$, and the mass $m_2$ of $A^0$ and $\tilde{\chi}_2^0$. The beam pipe is
defined to be along $\cos \theta = \pm 1$ and the center-of-mass
energy is $\sqrt s = 206$ GeV. The large difference between the IDM and
the MSSM models is due to the scalar \emph{versus} fermion nature of
the outgoing states.}
  \label{fig:3}
\end{figure*}

To illustrate observable outcomes we take help of the
event generator \textsc{MadGraph/MadEvent} \cite{Alwall:2007st},
which is to be further discussed in Section III B.
A comparison between the angular distributions of $e^+e^-\rightarrow
\tilde{\chi}_1^0 \tilde{\chi}_2^0$ and $e^+e^-\rightarrow H^0 A^0$
indicates that there are significant differences between the two
processes. Fig.~\ref{fig:3} shows the angular probability
distribution of produced $\tilde{\chi}_2^0$ and $A^0$ for
representative models with various neutralino and inert scalar
masses. As can be seen, $A^0$ is mainly produced with large
transverse momentum, while the $\tilde{\chi}_2^0$ distribution is
closer to isotropic.

\begin{figure*}[p]
  \includegraphics[height=58mm]{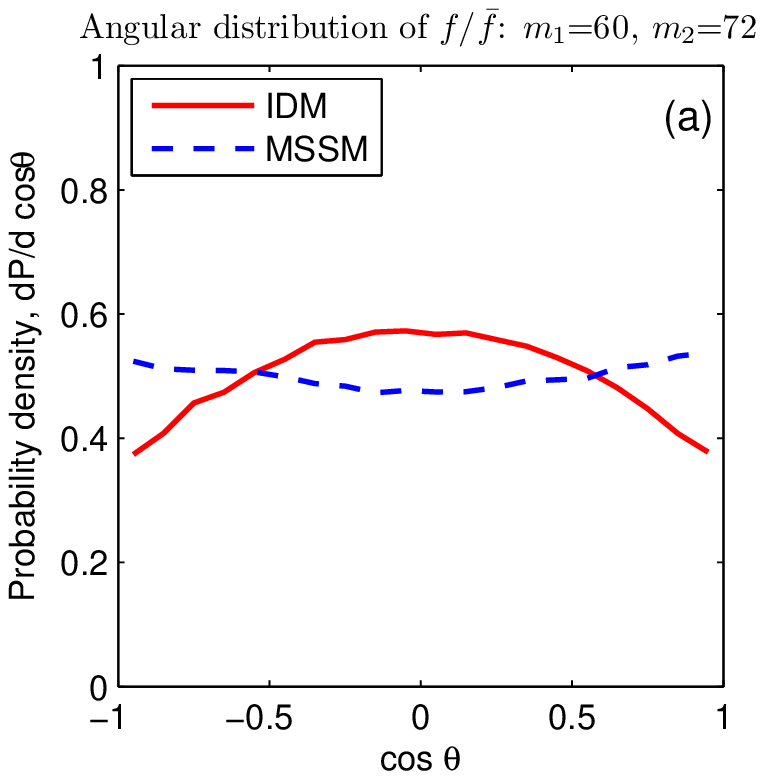}
  \includegraphics[height=58mm]{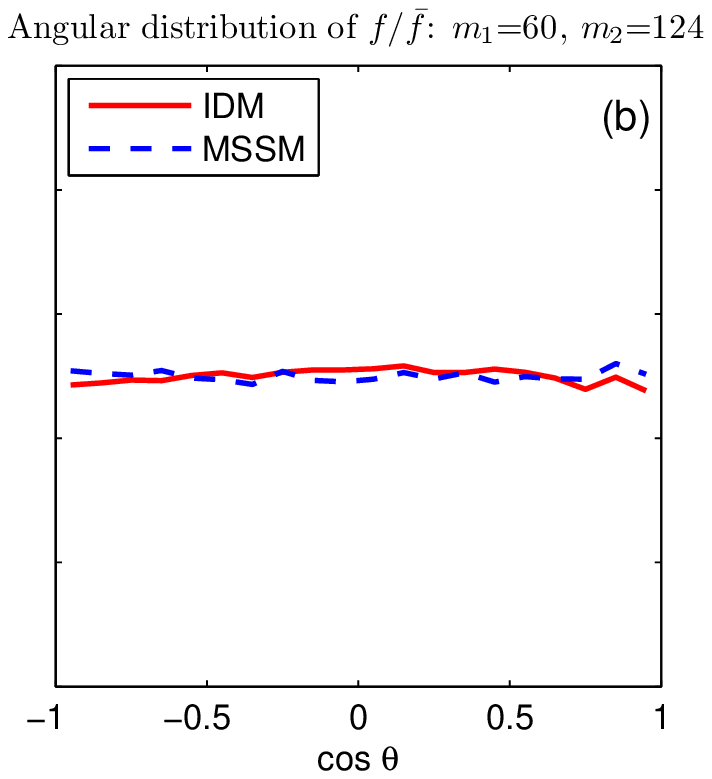}
  \includegraphics[height=58mm]{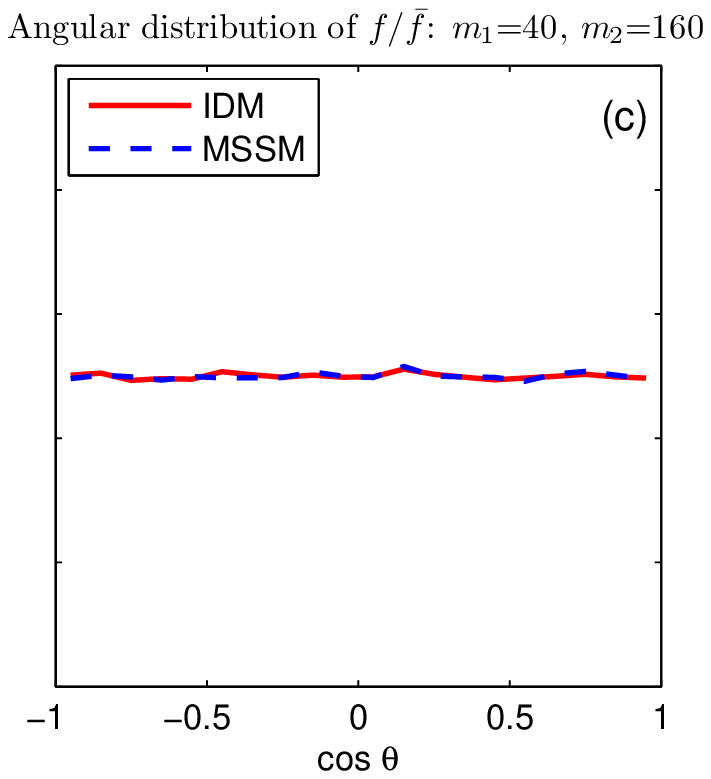}
\caption{Angular distribution of final state fermions from
$e^+e^-\rightarrow H^0 A^0\rightarrow H^0 H^0 f\bar{f}$ (IDM)
\emph{versus} $e^+e^-\rightarrow \tilde{\chi}_1^0 \tilde{\chi}_2^0$, $\tilde{\chi}_2^0\rightarrow \tilde{\chi}_1^0 f\bar{f}$ (MSSM). The models are the same as in Fig.~\ref{fig:3}. In (a) the velocity of the mother particle $A^0$/$\tilde{\chi}_2^0$ is large and the energy injected into the fermions during the decay is relatively low, and hence the discrepancy from Fig.~\ref{fig:3} can survive.}
  \label{fig:4}
\end{figure*}
\begin{figure*}[p]
  \includegraphics[height=58mm]{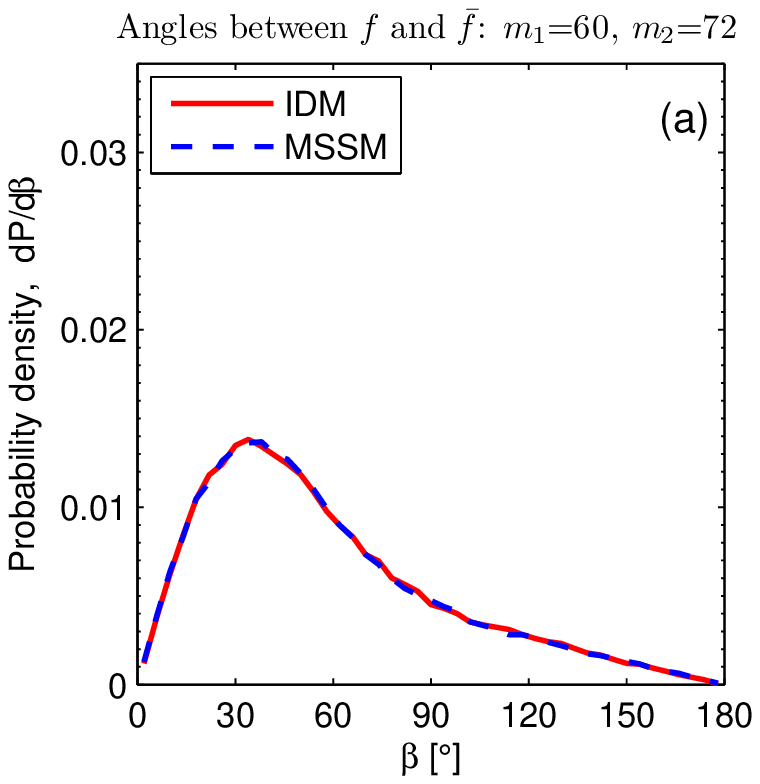}
  \includegraphics[height=58mm]{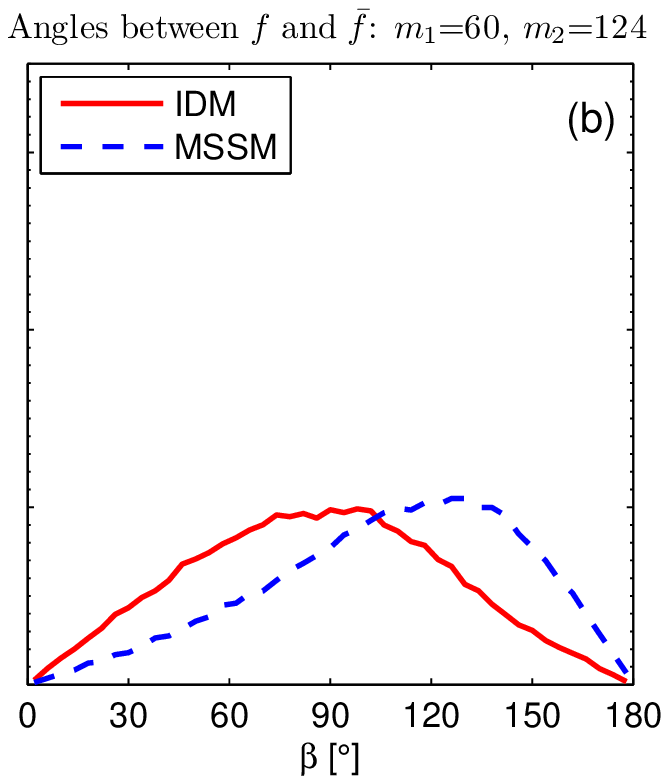}
  \includegraphics[height=58mm]{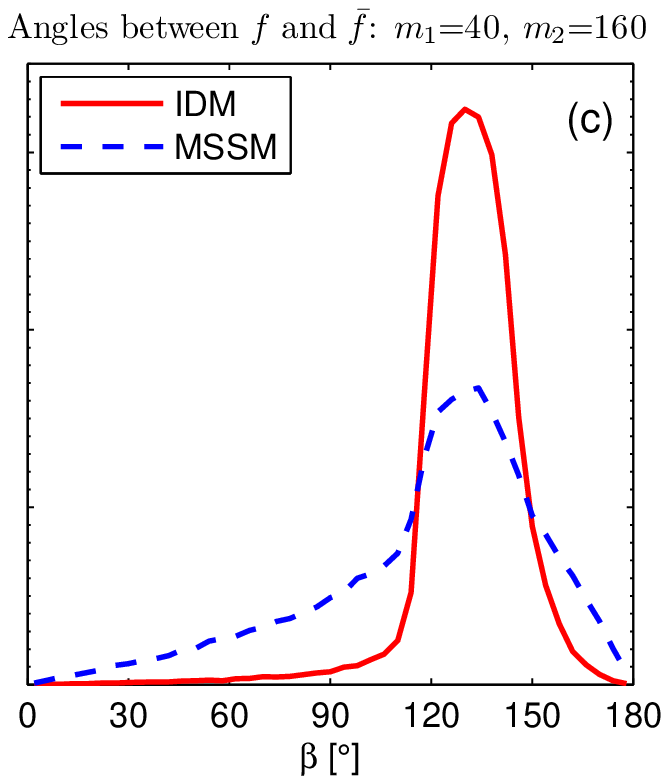}
\caption{Fermion opening angle distribution for the same processes and models as in
Fig.~\ref{fig:4}. $\beta$ is the angle between the two outgoing fermions $f$ and $\bar{f}$ as measured in the lab frame. In (a) the IDM and SUSY distributions
happen to be overlapping, but one should keep in mind that in the MSSM
the decay distribution can depend on the gaugino fraction of
$\tilde{\chi}_2^0$. Figs.~\ref{fig:3}-\ref{fig:5} are based on $10^5$ events per model, as generated with \textsc{MadGraph/MadEvent} \cite{Alwall:2007st}.}
  \label{fig:5}
\end{figure*}

Once including the decay processes one would expect the angular
distribution of final state particles to get a bit smeared out. In
order to properly include spin correlation and off-shell effects the
full $e^+e^-\rightarrow \tilde{\chi}_1^0\tilde{\chi}_2^0\rightarrow
\tilde{\chi}_1^0\tilde{\chi}_1^0 f\bar{f}$ and $e^+e^-\rightarrow
H^0 A^0\rightarrow H^0 H^0 f\bar{f}$ matrix elements, shown in
Figs.~\ref{fig:susy8}(e)-(h) and Fig.~\ref{fig:idm3}(c), have to be
calculated without factorizing them into production
($e^+e^-\rightarrow \tilde{\chi}_1^0\tilde{\chi}_2^0$,
$e^+e^-\rightarrow H^0 A^0$) and decay
($\tilde{\chi}_2^0\rightarrow\tilde{\chi}_1^0 f\bar{f}$,
$A^0\rightarrow H^0 f\bar{f}$) parts (see,
\textit{e.g.}, \cite{MoortgatPick:1997ny}). However, the analysis of
\cite{EspiritoSanto:2003} does not include any spin correlation
effects and we hence focus on the factorized $e^+e^-\rightarrow
\tilde{\chi}_1^0\tilde{\chi}_2^0$,
$\tilde{\chi}_2^0\rightarrow\tilde{\chi}_1^0 f\bar{f}$ process,
Figs.~\ref{fig:susy8}(a)-(d), for the neutralinos. Even if spin
correlations are absent for scalars, off-shell effects could in
principle make some difference and we hence use the unfactorized
process, Fig.~\ref{fig:idm3}(c), for the IDM.

Fig.~\ref{fig:4} shows the angular probability distribution of the
final state fermions for the same models as in Fig.~\ref{fig:3}.
Since we factorize the MSSM processes, and since the inert scalars
carry no spin, an isotropic decay in the rest frame of the mother
particle ($\tilde{\chi}_2^0$ or $A^0$) is expected. This makes the
final angular distributions considerably flatter, and the MSSM and
IDM models typically produce rather similar results, see
Figs.~\ref{fig:4}(b)-(c). However, if the velocity of the mother
particle is large and the energy injected into the $f\bar{f}$ pair
during the decay is relatively low, the boost can preserve some
significant difference. In other words, the MSSM and IDM final
fermion distributions will still differ for models where the
center-of-mass energy, $\sqrt{s}$, is well above the kinematical
limit for neutralino/inert scalar pair production, while at the same
time the mass difference between the produced particles is small,
see Fig.~\ref{fig:4}(a). Noteworthy, this is exactly the interesting
dark matter preferred region of rather light $H^0$
($m_{H^0}\!\!<\!m_W$) together with a considerable amount of
coannihilation with $A^0$ ($\Delta m\!\lesssim\! 10$ GeV)
\cite{Barbieri:2006dq,LopezHonorez:2006gr,Gustafsson:2007pc}.

It should be stressed that there are of course other observables
than individual scattering angles which may differ between the MSSM
and the IDM models. As an example Fig.~\ref{fig:5} shows the
$f\bar{f}$ opening angle distributions, where clear differences are
seen in Figs.~\ref{fig:5}(b)-(c). Which observables are important
depends on what cuts are imposed, and in general it can be hard to
extract any reliable information without doing a more complete
analysis. In any case, it should be clear from the above examples
that there exist differences between the MSSM and the IDM which
could hinder a direct application of the existing LEP II neutralino
production cross section limits on the inert scalars. We therefore
find it well motivated to study the LEP II bounds on the IDM more
closely.

\subsection{Method}

\subsubsection{General outline}
A DELPHI collaboration search for pair production of neutralinos in
the center-of-mass energy range 192--208 GeV
\cite{EspiritoSanto:2003} forms the basis of this paper. In
\cite{EspiritoSanto:2003} the absence of an excess above the
standard model background in the DELPHI data was used to put upper
limits on neutralino production cross sections. The analysis
included a careful standard model background generation and a
detailed detector simulation. The signal events were produced with
the help of \textsc{Susygen 2.2004} \cite{Katsanevas:1997fb}, and
\textsc{Jetset 7.4} \cite{Sjostrand:1993yb} was used for quark
fragmentation. Every event was tested against a number of searches
with cuts optimized for different final state topologies, each using
a sequential cuts approach. These searches, which are described in
detail in \cite{EspiritoSanto:2003}, were ordered according to 1)
acoplanar jets, 2) acoplanar leptons, 3) multijets, 4) multileptons,
and 5) asymmetric taus, and an event being accepted by one of the
selections was explicitly rejected by those appearing later in the
list. However, the searches were designed to be mutually exclusive
in order to substantially reduce the possibility of events being
capable of passing more than one of the searches. Based on the
resulting efficiencies for simulated signal events, 95\% C.L. upper
limits on neutralino production cross sections were derived and
presented as functions of neutralino masses in contour plots. Of
interest for this paper are Figs.~12 and 13 in
\cite{EspiritoSanto:2003} where $\tilde{\chi}_1^0\tilde{\chi}_2^0$
production followed by $\tilde{\chi}_2^0$ decay into
$\tilde{\chi}_1^0$ plus a fermion pair ($q\bar{q}$, $\mu^+\mu^-$,
$e^+e^-$, or according to the branching ratios of the $Z$ boson) was
assumed. Direct translation of these kinds of plots is what
previously has been used to estimate rough limits on the IDM.

The basic strategy of this work is simple: simulate
$e^+e^-\rightarrow H^0 A^0$ events, pass them through the same cuts
as those used in \cite{EspiritoSanto:2003} to determine IDM signal
efficiencies, and finally rescale the upper limits on production
cross sections derived in \cite{EspiritoSanto:2003}  in accordance
with the ratio between the MSSM and IDM efficiencies. However, in
reality the full analysis of \cite{EspiritoSanto:2003} is not
practically possible to reproduce in full detail. Instead we rely on
slightly different codes for event generation, a rough detector
simulation, and a somewhat simplified set of cuts. Thus a
reevaluation of also the MSSM efficiencies is required, both for
consistency and in order to check the reliability of our method.

\subsubsection{MSSM framework}
The free parameters of the R-parity conserving MSSM considered by \cite{EspiritoSanto:2003} are $\tan\beta$, $\mu$, $M_2$, $m_0$, $A$ and $m_A$. The gaugino mass parameters $M_1$ and $M_2$ are assumed to be related according to the mSUGRA unification relation: $M_1\!\approx\!$ 0.5$M_2$. By choosing a high common scalar mass parameter $m_0\!\sim\! 1$ TeV, \cite{EspiritoSanto:2003} put focus on models with sfermions much heavier than the produced neutralinos. For fixed values of $\tan\beta$ (\textit{i.e.} the ratio between the vacuum expectation values of the two Higgs doublets), \cite{EspiritoSanto:2003} performed scans over $M_2$ and the Higgs mass parameter $\mu$ to find models with suitable values of the $\tilde{\chi}_1^0$ and $\tilde{\chi}_2^0$ masses. The common trilinear coupling $A$ and the pseudoscalar Higgs mass $m_A$ typically have no noticeable influence on the results.

Because of the assumed unification relation between $M_1$ and $M_2$,
there are no models with the mass of $\tilde{\chi}_2^0$ much larger
than twice the mass of $\tilde{\chi}_1^0$. Therefore
\cite{EspiritoSanto:2003} considered $\tilde{\chi}_1^0
\tilde{\chi}_3^0$ production followed by the decay
$\tilde{\chi}_3^0\rightarrow\tilde{\chi}_1^0 f\bar{f}$ when they
explored larger neutralino mass differences. However, their notation
was to not explicitly write out any $\tilde{\chi}_3^0$, but instead
use $\tilde{\chi}_2^0$ for any heavier neutralino. We follow the
same convention here. Although one can in general find several
points within the ($\tan\beta$,$M_2$,$\mu$) parameter space giving
the same set of ($\tilde{\chi}_1^0$,$\tilde{\chi}_2^0$) masses, the
differences in efficiencies are typically small.

\subsubsection{Signal generation}
For the generation of signal events we use the
\textsc{MadGraph/MadEvent} package \cite{Alwall:2007st}. There are
two main reasons for this choice: 1) It is possible to generate both
MSSM and IDM events with \textsc{MadGraph/MadEvent}. 2) The code
allows multiparticle final states, giving the opportunity to
generate the complete $2\rightarrow 4$ processes in one step, thus
fully taking spin correlation and off-shell effects into account. In
agreement with \cite{EspiritoSanto:2003}, the MSSM processes are
factorized into $e^+e^-\rightarrow
\tilde{\chi}_1^0\tilde{\chi}_2^0$, followed by either
$\tilde{\chi}_2^0\rightarrow \tilde{\chi}_1^0 q\bar{q}$,
$\tilde{\chi}_2^0\rightarrow \tilde{\chi}_1^0 \mu^+\mu^-$, or
$\tilde{\chi}_2^0\rightarrow \tilde{\chi}_1^0 e^+e^-$. The decays
are arranged to be isotropic in the rest frame of
$\tilde{\chi}_2^0$, after which the decay products are boosted into
the lab frame. The IDM events, on the other hand, are generated via
the unfactorized processes $e^+e^-\rightarrow H^0 A^0\rightarrow H^0
H^0 f\bar{f}$, where $f\bar{f}$ represents either $q\bar{q}$,
$\mu^+\mu^-$, or $e^+e^-$. As a consistency check we also study
numerous unfactorized MSSM processes as well as factorized IDM
processes. Although slight deviations in for example the angular
distribution of produced particles can be found between the
factorized and unfactorized MSSM processes, the discrepancies left
at the level of efficiency determination are quite small. For the
IDM this agreement is even better.

We fix the center-of-mass energy $\sqrt{s}$ to 206 GeV in our
analysis, and note that the final results hardly depend on the exact
value.

\textsc{Pythia 6.4} \cite{Sjostrand:2006za} is used for the
fragmentation and hadronization of final state quarks, and jets are
reconstructed with the PYCLUS algorithm. The minimum transverse jet
separation is set to $d_{join}=10$ GeV, while at the same time
requiring at least two reconstructed jets, all in agreement with
\cite{EspiritoSanto:2003}.

\subsubsection{Cuts}

Different cuts are imposed depending on whether the final state
includes quark or lepton pairs. We mimic the cuts of the acoplanar
jets search and the acoplanar leptons search in
\cite{EspiritoSanto:2003} as closely as possible, and apply them to
the quark and lepton events, respectively. Muon and electron identifications are performed with the help of
tabulated acceptances extracted from an earlier detailed study of
the DELPHI detector \cite{Klas}.
Given the four-momentum of the lepton, a probability for
identification is assigned and used to generate identification
status.

Because of \textit{e.g.}
our lack of a detailed detector simulation, some of the cuts have to
be somewhat simplified, and some are not even possible to carry out.
However, much effort has been put into accounting for all major effects in order to
assure the reliability of our final conclusions.

In principle an event not passing the acoplanar jets or acoplanar
leptons cuts could instead happen to be accepted by one of the
subsequent searches (\textit{i.e.}~the multijets, multileptons, or
asymmetric taus selections). Owing to the design of the searches we
expect these events to be quite rare, and that the relative MSSM and
IDM efficiencies in any case can be fairly represented by the
corresponding results from the acoplanar searches. We therefore only
impose one set of cuts on each event: the acoplanar jets (for
$q\bar{q}$ final states) or the acoplanar leptons (for $\mu^+\mu^-$
and $e^+e^-$ final states) selections. The exact selection cuts are
described in appendices A and B.

\subsubsection{Calculating efficiencies}
We start our analysis by generating (on average a handful of
different) MSSM models for each set of
($\tilde{\chi}_1^0$,$\tilde{\chi}_2^0$) masses considered. We create
the models using the spectrum generator \textsc{Suspect 2.34}
\cite{Djouadi:2002ze}, and feed them into \textsc{MadGraph/MadEvent}
to simulate MSSM events. In total we generate more than
$15\times10^6$ $\tilde{\chi}_1^0\tilde{\chi}_2^0$ events distributed
among roughly 200 different MSSM models. Efficiencies are then
calculated by passing the generated events through our cuts. As a
reliability check of our analysis, the found efficiencies are
compared with those of Table 8 in \cite{EspiritoSanto:2003}.
Although an exact agreement is too much to demand, the ratios
between our calculated efficiencies and those in
\cite{EspiritoSanto:2003} should at least stay rather constant for
all ($\tilde{\chi}_1^0$,$\tilde{\chi}_2^0$) mass points. Indeed we
find that this ratio stays in the interval 0.7--1.0 for most models,
a result accurate enough for our needs. Some larger deviations
occur, but this happens mainly in not very important regions of the
parameter space, and can possibly be understood by taking into
account the uncertainties in the jet energy determination by the
true detector. See Section V for further discussion.

Within the IDM the cross sections for production of $H^0 A^0$ events
are on the other hand completely specified once $m_{H^0}$ and
$m_{A^0}$ are given, and hence only one model for each set of
($H^0$,$A^0$) masses needs to be examined. Again
\textsc{MadGraph/MadEvent} is used to simulate (about $4\times10^6$)
signal events, which are subsequently taken through the same cuts in
order to extract the corresponding IDM efficiencies.

\subsubsection{Constraining the IDM}
We can now determine the ratio between our calculated MSSM and IDM
efficiencies as a function of the produced neutralino/inert scalar
masses. These are then used to rescale the cross section upper limits presented in \cite{EspiritoSanto:2003}.
It should be noted that although the 14 mass
combinations given in Table 8 in \cite{EspiritoSanto:2003}
serve as our starting check points, we investigate many more to
cover the complete ($m_{H^0}$,$m_{A^0}$) plane and to be more
detailed around more critical regions.

In the decay of $A^0$, fermion pairs are produced at tree level via
s-channel $Z$ bosons, and the branching ratios into fermions are
therefore expected to be close to those of an on-shell $Z$ boson. It
is therefore natural to focus on the cross section limits provided
by Fig.~13(d) in \cite{EspiritoSanto:2003}, where the final fermions
are assumed to be produced in accordance with the branching ratios
of the $Z$ boson. This is clearly appropriate for models with a
large mass splitting $\Delta m$ (and specifically when $\Delta m\geq
m_Z$ so that $Z$ can be on-shell), but some modifications may be
needed for models where the intermediate $Z$ boson is very virtual
and the branching may be somewhat different. In other words, when
$\Delta m$ is small we need to evaluate the branching ratios to find
out whether some modifications might be needed (this calculation is
done with \textsc{MadGraph/MadEvent}).

Finally, we calculate the $H^0 A^0$ production cross section as a function of
$m_{H^0}$ and $m_{A^0}$, and compare it with our derived cross section upper limits in order to constrain the IDM parameter space.

\section{Results}
\begin{figure}[!t]
    \centering
        \includegraphics[width=1.0\columnwidth]{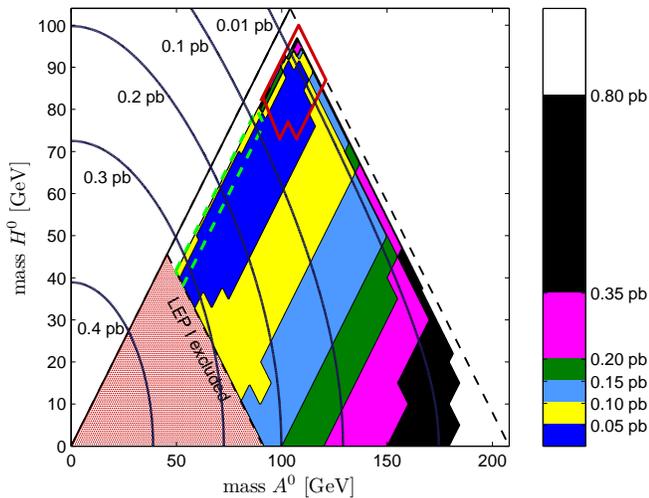}
\caption{Production cross section upper limits as extracted from
Fig.~13(d) in \cite{EspiritoSanto:2003}. For models inside the (red)
solid [(green) dashed] contour the limits are rescaled by a factor
0.9 (1.1) before being applied to $H^0A^0$ production. The solid
(dark blue) contour lines indicate the $e^+e^-\rightarrow H^0A^0$
cross section. The (red) dotted-shaded region, where
$m_{H^0}\!+m_{A^0}\!<\!m_Z$, is excluded by LEP I data on the Z
boson width. The upper right dashed line shows the LEP II
kinematical limit.}
    \label{fig:idm_LEP_limits}
\end{figure}

Under our imposed cuts the resulting IDM and MSSM
efficiencies turn out to be quite similar, an appealing, although
not at all trivial, result.

The efficiencies are first determined for each individual
channel ($q\bar{q}$, $\mu^+\mu^-$, $e^+e^-$), after which those are
combined into an efficiency representing the actual branching
ratio. This combination is done by weighting the channels in
accordance with the decay branching ratios of the Z boson
(\textit{i.e.} the $q\bar{q}$ efficiency is given the highest
weight).

In general we observe that the ratio between our derived IDM and MSSM efficiencies is quite insensitive to the very details of the imposed cuts, and we estimate our sensitivity in determining this ratio to be of the order of 10 \!\%.

We find that whenever $m_{H^0}\!\lesssim 80$ GeV the IDM efficiencies typically are a few percent higher than those of the corresponding MSSM models. An important observation is that we find no mass combinations in this region where the MSSM gives a higher efficiency than the IDM, and it is therefore appropriate to apply at least as hard production cross section upper limits on the inert scalars as those put on the neutralinos in \cite{EspiritoSanto:2003}.

In the specific region defined by 8 GeV \!$<\Delta m<$\! 15 GeV and
$m_{H^0}\!\lesssim 85$ GeV, the IDM efficiencies are found to be
about a factor 1.15-1.20 higher than those of the MSSM. On noting
that the models with the lowest $\Delta m$ have a slightly higher
branching into neutrinos compared to ordinary Z boson decay, we in
this region adopt a conservative factor of 0.9 with which we rescale
the neutralino production limits given in Fig.~13(d) in
\cite{EspiritoSanto:2003}. This region is encircled with a (green)
dashed line in Fig.~\ref{fig:idm_LEP_limits}.

Among the remaining $m_{H^0}\!\gtrsim 80$ GeV models we find some
for which the ratio between the IDM and MSSM efficiencies drops down
to 0.9. We therefore use a factor of 1.1 for the rescaling here, and
this region is encircled with a red solid line in
Fig.~\ref{fig:idm_LEP_limits}.

Except for in the low $\Delta m$ and high $m_{H^0}$ regions
mentioned above we find it appropriate to apply the same production
limits as for the neutralinos. While this might be argued to be too
conservative, the points where harder limits could possibly be
imposed are anyway far from excluding any IDM model.

By utilizing the limits on the $\tilde{\chi}_1^0 \tilde{\chi}_2^0$
production from Fig.~13(d) in \cite{EspiritoSanto:2003} we find,
after rescaling, upper limits on the $H^0 A^0$ production cross
section as a function of $m_{H^0}$ and $m_{A^0}$. The cross section
limits, and the regions where we impose rescaling, are found in
Fig.~\ref{fig:idm_LEP_limits}.
\begin{figure}[!t]
    \centering
        \includegraphics[width=0.8\columnwidth]{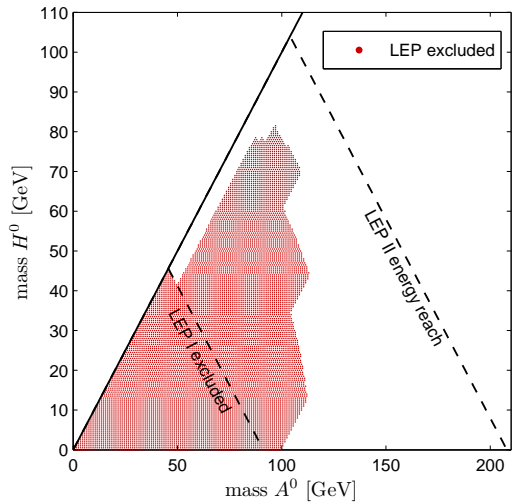}
\caption{LEP exclusion plot. The (red) dotted-shaded region
indicates the region of the ($m_{H^0}$,$m_{A^0}$) plane excluded by
LEP data. The lower left triangle, where
$m_{H^0}\!+m_{A^0}\!<\!m_Z$, is excluded by LEP I data on the Z
boson width. The remaining part of the shaded region is excluded by
our LEP II analysis. Shown is also the LEP II kinematical limit.
Since we are assuming $m_{H^0}\!<\!m_{A^0}$ the upper left region is
not accessible.}
    \label{fig:idm_LEP_excl}
\end{figure}
Comparing these with the calculated $e^+ e^- \rightarrow H^0 A^0$
cross sections, which also are shown in
Fig.~\ref{fig:idm_LEP_limits}, finally tells us which IDM models are
excluded.

The resulting exclusion plot is shown in
Fig.~\ref{fig:idm_LEP_excl}. Roughly speaking, our LEP II analysis
exclude models satisfying $m_{H^0}\!<\! 80$ GeV, $m_{A^0}\!<\!100$
GeV and $\Delta m\!>\!8$ GeV. The sharp transition at $\Delta
m\!\!=\!\!8$ GeV comes from the steep gradient of the cross section
upper limit present in Fig.~13(d) in \cite{EspiritoSanto:2003}.

In \cite{Barbieri:2006dq} it has previously been estimated that the $H^0 A^0$ production cross section is below the existing LEP II upper limits for models with $m_{H^0}\approx 70$ GeV and $\Delta m\lesssim 10$ GeV. We note that our results put somewhat harder constraints in that region. Moreover, \cite{Pierce:2007ut} presented $m_{H^0}+m_{A^0}>130$ GeV as a rule of thumb for models to be consistent with the LEP II data. For $m_{H^0}>30$ GeV our analysis hence imposes harder constraints than that.
\begin{figure}[t]
    \centering
        \includegraphics[width=0.94\columnwidth]{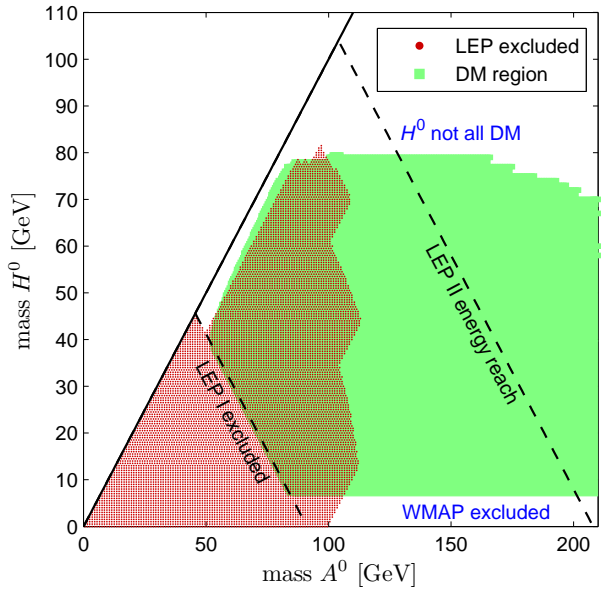}
        \caption{LEP II limits and dark matter, $m_h\!=\!200$ GeV. The LEP
excluded region of Fig.~\ref{fig:idm_LEP_excl} together with a
(green) gray region indicating points in the ($m_{H^0}$,$m_{A^0}$)
plane where there exist models with $m_h\!=\!200$ GeV capable of
providing a good dark matter (DM) candidate $H^0$ with relic density
in agreement with WMAP. Models above the (green) gray region are
still allowed but cannot account for all dark matter. Models below
the (green) gray region leave a relic density higher than the WMAP
upper limit and are hence ruled out.}
    \label{fig:idm2b_excl_DM}
\end{figure}

\begin{figure}[t]
    \centering
        \includegraphics[width=0.94\columnwidth]{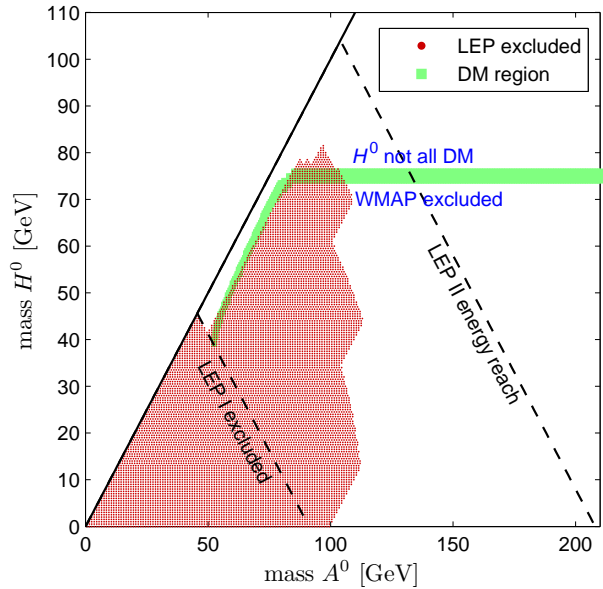}
\caption{LEP II limits and dark matter, $m_h\!=\!500$ GeV. The LEP
excluded region of Fig.~\ref{fig:idm_LEP_excl} together with a
(green) gray band indicating points in the ($m_{H^0}$,$m_{A^0}$)
plane where there exist models with $m_h\!=\!500$ GeV capable of
providing a good dark matter (DM) candidate $H^0$ with relic density
in agreement with WMAP. Models above the (green) gray band are still
allowed but cannot account for all dark matter. Models below the
(green) gray band all leave a relic density higher than the WMAP
upper limit and are hence ruled out.}
    \label{fig:idm3b_excl_DM}
\end{figure}

In order to find out what consequences this LEP II constraint has
for $H^0$ as a dark matter candidate we perform an extensive scan
over the IDM parameter space using \textsc{DarkSusy} \cite{ds}
interfaced with \textsc{FormCalc} \cite{FormCalc}. The relic density
is calculated, including coannihilations with $A^0$ and $H^{\pm}$,
for models respecting Eqs.~(\ref{eq:vacuumstability}),
(\ref{eq:perturbativity}), (\ref{eq:EWPT}) and (\ref{eq:Zwidth}). We
also include the limit $m_{H^{\pm}}\!\!>70$ GeV, which is based on
direct translation of the limits from LEP II chargino search results
\cite{Pierce:2007ut}.

Fig.~\ref{fig:idm2b_excl_DM} and
Fig.~\ref{fig:idm3b_excl_DM} show, for  $m_h\!=\!200$ GeV and
$m_h\!=\!500$ GeV, respectively, the regions where $H^0$ can have the
right relic abundance to account for all dark matter.

Because the only existing (tree level) $H^0$ self-annihilation
channel into fermions is via s-channel $h$ exchange, models with a
heavy $h$ can only provide a relic density in agreement with the
WMAP data if either: 1) $\Delta m$ is small so that coannihilation
with $A^0$ becomes important, or 2) $m_{H^0}\!\sim\! 75$ GeV so that
the efficient annihilations into massive gauge bosons are of just
about the right magnitude in the early Universe. If the
coannihilations are too strong, or if $m_{H^0}$ is too high, the
relic density will fall below the WMAP lower limit, in which case
the model is still allowed but cannot explain all dark matter. These
regions are easily identified in Fig.~\ref{fig:idm3b_excl_DM}.

In Fig.~\ref{fig:idm2b_excl_DM} the mass of $h$ is small
enough to allow the $H^0$ self-annihilating cross section into fermions to be
large enough to obtain a correct relic density in a larger part of
the ($m_{H^0}$,$m_{A^0}$) plane. The smaller value of $m_h$ also
allows, because of cancellations between the diagrams contributing
to annihilations into massive gauge bosons, for a slightly heavier dark matter
candidate $H^0$.

As can be seen in Figs.~\ref{fig:idm2b_excl_DM} and
Fig.~\ref{fig:idm3b_excl_DM} the LEP II data can exclude many IDM
models of interest from a dark matter perspective, although
definitely not all of them.

\section{Discussion}

Although we find that appropriate upper limits to impose on the
production cross section for inert scalars are similar to those
already derived for neutralinos in \cite{EspiritoSanto:2003}, this
result is \textit{a priori} far from trivial owing to the different
properties of the particles.

It is encouraging that the effect of including realistic
selection efficiencies is not to invalidate, but rather to increase the
confidence in simple applications of limits from other MSSM searches
directly onto the IDM. However, one must keep in mind that rather
different sets of cuts may be used in other searches, and hence any
robust prediction requires an evaluation of the corresponding IDM
efficiencies.

One may worry that some of the simplifications used in this
analysis misses some important effects. For example we do not
include any initial state radiation from the incoming $e^+e^-$
pair. While this is certainly important for the models excluded by
LEP I which satisfy $m_{H^0}\!+\!m_{A^0}\!\!<\!m_Z$ (and for which
the intermediate s-channel Z bosons can be put on-shell by radiating
off photons from the incoming particles) we do not expect the
initial state radiation to be very important for our mass range of
interest.

Another concern is that the detector generally fails to detect some
amount of the energy in high energy jets. On assuming a perfect jet
energy determination we indeed find that for models with $\Delta
m\!\gtrsim\! 100$ GeV our derived MSSM efficiencies do not match
those of \cite{EspiritoSanto:2003} as well as they do for lower
$\Delta m$ models. However, by artificially scaling down the jet
energy of each event with on average 10\%-20\% before applying the
cuts we find good agreements also for these models. An important
observation to keep in mind is that we find the ratios between the
IDM and MSSM efficiencies to be practically insensitive to the
downscaling magnitude. Therefore our final IDM limits do not depend
on the very details of the true energy losses. Also, the $H^0 A^0$
production cross section is in general too low within the high
$\Delta m$ region for any models to be excluded anyway.

In \cite{EspiritoSanto:2003} the acoplanar searches were
each divided into four subselections (these subselections are the
last 1)-4) cuts presented in each of appendices A and B). Within their analysis these were treated as
independent selections, and we should hence actually check the
efficiencies under each of them separately when trying to extract
the IDM limits.
However, the ratios between our derived IDM and MSSM
subselection efficiencies are found to be practically equal to the values for the corresponding complete acoplanar selections. We hence
conclude that this does not constitute a problem for our method.

While our analysis certainly includes some approximations, we have
consistently been quite conservative in order to minimize the risk
of overestimating the final constraints on the IDM. Hence,
our derived LEP II limits on the inert scalar masses should be
robust.

\section{Summary}
In this paper we have investigated what limits can be inferred on
the IDM from an already existing LEP II neutralino search. In
performing this translation we have used a method respecting the
differences between inert scalars and neutralinos. We have shown
that the LEP II data can exclude a significant part of the parameter
space, but also that many models providing a good WIMP dark matter
candidate are still valid.

\section*{Acknowledgments}
We thank Klas Hultqvist for numerous fruitful discussions regarding
the analysis of \cite{EspiritoSanto:2003} as well as the DELPHI
detector in general, for providing us with the lepton identification
code, and for careful reading of the manuscript. Thanks also to Anna
Lipniacka and Per Johansson for useful comments. E.L. would like to
thank Johan Alwall for valuable technical assistance with
\textsc{MadGraph/MadEvent}. M.G.~is supported by the INFN and the EU
FP6 Marie Curie Research \& Training Network ``UniverseNet''
(MRTN-CT-2006-035863). J.E. thanks the Swedish Research Council (VR)
for funding support.

\section*{Appendix A}
Below follows the acoplanar jets selection cuts we imposed on the
quark final states after hadronization in \textsc{Pythia} and
subsequent rejection of particles lost in the beam pipe (which is
taken to cover a polar angle of $10^\circ$). A particle was assumed
to carry time projection chamber (TPC) information if and only if it
was more than $25^\circ$ away from the beam axis. (See
\cite{Abreu:1995uz} for information about the TPC pad rows in the
DELPHI detector.) Although some simplifications have been
unavoidable, most of our cuts are very close to the original ones.
For a list of original cuts see \cite{EspiritoSanto:2003}.
\begin{itemize}
    \item A minimum of two charged particles, with at least one of them having a transverse momentum above 1.5 GeV, and a total transverse energy (defined as the sum of the absolute values of the transverse momenta of all individual particles) above 4 GeV, was required.
    \item At least five charged particles had to carry TPC information.
    \item The scalar sum of momenta of the particles carrying TPC information had to exceed both 4 GeV and 10\% of the total jet energy.
    \item Exactly two jets were required.
    \item All jets had to have a polar angle above $10^\circ$.
    \item Each jet had to contain at least one particle with TPC information.
    \item The particles emitted within $30^\circ$ of the beam axis had to carry less than 60\% of the total jet energy.
    \item The absolute value of the cosine of the polar angle of the total momentum had to be smaller than 0.9.
    \item The total transverse momentum (defined as the norm of the vector sum of the individual transverse momenta) had to be larger than 6 GeV.
    \item The momentum of a jet divided by its energy had to exceed 0.5 for the most energetic jet and 0.4 for the second most energetic jet.
    \item The total transverse momentum had to be larger than 15 GeV, or the acollinearity (defined as the supplement of the angle between the jets) had to exceed $40^\circ$, or the ratio between the momentum and the energy of the second most energetic jet had to be above 0.8.
    \item If the total transverse momentum was below 12 GeV, the average momentum of the particles with TPC information had to lie between 0.8 GeV and 8 GeV, and the energy of the most energetic neutral particle had to be below 40\% of the total jet energy (if the total jet energy exceeded 20 GeV) or below 5 GeV (if the total jet energy was lower than 20 GeV).
    \item If the total jet energy was below 50 GeV, the average momentum of the particles with TPC information had to lie between 0.8 GeV and 8 GeV, and the momentum of the most energetic charged particles had to lie between 5\% and 70\% of the total jet energy.
    \item If the total jet energy was below 20 GeV, the average momentum of the particles with TPC information had to lie between 0.8 GeV and 4 GeV but at least be larger than 20\% of the total jet energy, the momentum of the most energetic charged particle had to lie between 10\% and 60\% of the total jet energy, and the energy of the most energetic neutral particle had to be below 35\% of the total jet energy.
    \item If there were any neutral particles, the norm of the vector sum of the jet momenta and the momentum of the most energetic neutral particle had to exceed 2.5 GeV.
    \item All neutral particles had to have energies less than 60 GeV.
    \item All charged particles had to have energies less than 20 GeV.
    \item The jets had to pass one of the following cuts:
\begin{enumerate}
    \item The invariant mass had to be below 10\% of $\sqrt{s}$, the missing mass had to be above 70\% of $\sqrt{s}$, the total transverse momentum had to be above 7 GeV, and the scaled acoplanarity (defined as the sine of the minimum angle between a jet and the beam axis times the supplement of the angle between the projections of the jets onto a plane perpendicular to the beam axis) had to exceed $40^\circ$.
    \item The invariant mass had to lie between 10\% and 30\% of $\sqrt{s}$, the missing mass had to be above 60\% of $\sqrt{s}$, the total transverse momentum had to be above 8 GeV, and the scaled acoplanarity had to exceed $25^\circ$.
    \item The invariant mass had to lie between 30\% and 50\% of $\sqrt{s}$, the missing mass had to be above 45\% of $\sqrt{s}$, the total transverse momentum had to lie between 12 GeV and 35 GeV, the total longitudinal momentum (defined as the norm of the vector sum of the individual longitudinal momenta) had to be below 35 GeV, the scaled acoplanarity had to exceed $25^\circ$, and the acollinearity had to be lower than $55^\circ$.
    \item The invariant mass had to lie between 50\% and 70\% of $\sqrt{s}$, the missing mass had to be above 20\% of $\sqrt{s}$, the total transverse momentum had to lie between 12 GeV and 35 GeV, the total longitudinal momentum had to be below 35 GeV, the scaled acoplanarity had to exceed $25^\circ$, and the acollinearity had to be lower than $55^\circ$.
\end{enumerate}
\end{itemize}

\section*{Appendix B}
Below follows the acoplanar leptons selection cuts we imposed on the
lepton final states. A particle was assumed to carry TPC information
if and only if it was more than $25^\circ$ away from the beam axis.
(See \cite{Abreu:1995uz} for information about the TPC pad rows in
the DELPHI detector.) Although some simplifications have been
unavoidable, most of our cuts are very close to the original
ones. For a list of original cuts see \cite{EspiritoSanto:2003}.
\begin{itemize}
  \item Two leptons needed to be identified.
    \item At least one of the leptons had to have a transverse momentum above 1.5 GeV, and a total transverse energy above 4 GeV was required.
  \item Each lepton had to have an energy above 1 GeV.
  \item Each lepton had to carry TPC information.
  \item The acoplanarity between the leptons had to be larger than $10^\circ$.
  \item The acollinearity between the leptons had to be larger than $10^\circ$.
  \item The absolute value of the cosine of the polar angle of the total momentum had to be smaller than 0.9.
    \item The total transverse momentum had to be larger than 6 GeV.
    \item The leptons emitted within $30^\circ$ of the beam axis had to carry less than 70\% of the total lepton energy.
    \item If the total transverse energy was below 100 GeV and the total missing momentum was above 45 GeV, then the charge times the cosine of the polar angle of the most energetic lepton was required to be positive.
    \item If the momentum of the most energetic lepton was between 40 GeV and 100 GeV and the total energy within $15^\circ$ of this lepton was between 40 GeV and 100 GeV, then the charge times the cosine of the polar angle was required to be greater than -0.65 for both leptons.
  \item The leptons had to pass one of the following cuts:
\begin{enumerate}
    \item The invariant mass had to be below 10\% of $\sqrt{s}$, the missing mass had to be above 70\% of $\sqrt{s}$, the total transverse momentum had to be above 7 GeV, and the acoplanarity (defined as the supplement of the angle between the projections of the leptons onto a plane perpendicular to the beam axis) had to exceed $40^\circ$.
    \item The invariant mass had to lie between 10\% and 30\% of $\sqrt{s}$, the missing mass had to be above 45\% of $\sqrt{s}$, the total transverse momentum had to be above 10 GeV, and the acoplanarity had to exceed $25^\circ$.
    \item The invariant mass had to lie between 30\% and 55\% of $\sqrt{s}$, the missing mass had to be above 20\% of $\sqrt{s}$, the total transverse momentum had be above 12 GeV, and the acoplanarity had to exceed $15^\circ$.
    \item The invariant mass had to lie between 55\% and 70\% of $\sqrt{s}$, the missing mass had to be above 20\% of $\sqrt{s}$, the total transverse momentum had to be above 12 GeV, and the acoplanarity had to exceed $15^\circ$.
\end{enumerate}
\end{itemize}


\end{document}